\documentclass{elsart3p}
\usepackage{graphicx}
\journal{NIM A}
\begin{document}
\begin{frontmatter}
\title{Application of introduced nano-diamonds for the study
\\ of carbon
condensation during detonation of high explosives}
\author[igl]{K.\,A. Ten\corauthref{cor}},
\corauth[cor]{Corresponding author. Tel. : +7 9139031515;\\
fax.:~+7 383 3331612}
\ead{ten@hydro.nsc.ru}
\author[binp]{V.\,M. Aulchenko},
\author[igl]{L.\,A. Lukjanchikov},
\author[igl]{E.\,R. Pruuel},
\author[binp]{L.\,I. Shekhtman},\ead{l.i.shekhtman@inp.nsk.su}
~\author[che]{B.\,P. Tolochko},
\author[che]{I.\,L. Zhogin},
\ead{i.l.zhogin@inp.nsk.su}
\author[binp]{V.\,V. Zhulanov}

\address[igl]{Lavrentjev Institute for Hydrodynamics SB RAS,
  Lavrentjev ave. 15,
 630090 Novosibirsk, Russian Federation}
\address[che]{Institute of Solid State Chemistry SB RAS,
 Kutateladze str. 18,
630128 Novosibirsk, Russian Federation}
\address[binp]{Budker Institute of Nuclear Physics SB RAS,
  Lavrentjev ave. 11,
 630090 Novosibirsk, Russian Federation}

\begin{abstract}
This paper describes experimental studies of the formation of
nano-diamonds during detonation of TNT/RDX 50/50 mixture with
small-angle x-ray scattering (SAXS) method at a synchrotron radiation
beam on VEPP-3 accelerator. A new experimental method with
introduction of nano-diamonds into the explosive has been
applied. Inclusion of the diamonds obtained after detonation
into the TNT and RDX explosives allows modelling of  the case of
instant creation of nano-diamonds during detonation.
\end{abstract}
\end{frontmatter}

\section{Introduction}
Although  quite large number of publications discusses a
synthesis of nano-diamonds during an explosion, the question of
its formation and even more common problem of carbon
condensation  at detonation of an explosive with negative oxygen
balance is being under discussion till the present time. An
answer to this question is important for both the understanding
of physics of the phenomenon and for the estimation of energy
which is generated during exothermal coagulation of carbon
clusters.

The first direct experimental studies of growth of an exploded
nano-diamonds became possible due to application of synchrotron
radiation, namely the possibility of detection of photons after
small-angle diffraction with a frequency up to 4 MHz. The
results obtained demonstrated that the intensity of SAXS signal
started
to grow from zero at the detonation front and its growth
continued during several microseconds [1--2].

A hypothesis suggested based on this result, that a creation of
carbon particles (including nano-diamonds) takes place beyond
the chemical reaction zone (in this case its size is around 0.7
mm). However this result can be used as well as confirmation of
instantaneous creation of exploded nano-diamonds.   One has to
suppose for that the low value of SAXS intensity at the
detonation front is due to small \emph{contrast} of
nano-diamonds. In SAXS method the intensity of scattered signal
at the first approximation is proportional to squared difference
of densities of the scattering particle and the medium
(detonation products -- DP). At DP density in the beginning of
about 2 g/cm${}^3$ the signal can be too small to be detected
and only during DP decay, i.e.\ when the density is reduced, it
can be registered. This can also explain a tendency for the
increase of the zone of the signal maximum delay with the
diameter of the explosive [2], as DP decay will take more time
in such case.

In the present work the study of the starting point of
nano-diamonds formation is performed during detonation of the
mixture of  TNT/RDX 50/50 with SAXS method at the synchrotron
radiation (SR) beam of VEPP-3 accelerator. A new experimental
arrangement is applied with nano-diamonds introduced beforehand
into the charge of high explosive (HE). Known initial presence
of nano-diamonds in pressed HE allows to determine the dynamics
of  exploded nano-diamonds growth with better precision.

\section{Experimental technique}
Experimental work has been performed in Budker INP (at VEPP-3,
wiggler with 2T field, average X-ray energy of 20 keV, bunch
frequency of 4-8 MHz and bunch duration of about 1 ns).
Electronic circuit of the detector allows recording of 32 frames
with the position distribution of scattered photons  performed
every 500, 250 or 125 ns. In the last case the storage ring
accelerates two bunches rotating at the opposite points at the
orbit. As the average current of the accelerator have to stay
the same, this arrangement leads to the decrease of the signal
to noise ratio (the dispersion of experimental points is
larger), but allows improvement of time resolution.

\begin{figure}[h]
\centering
\includegraphics*[bb=1 1 69 43,width=80mm]{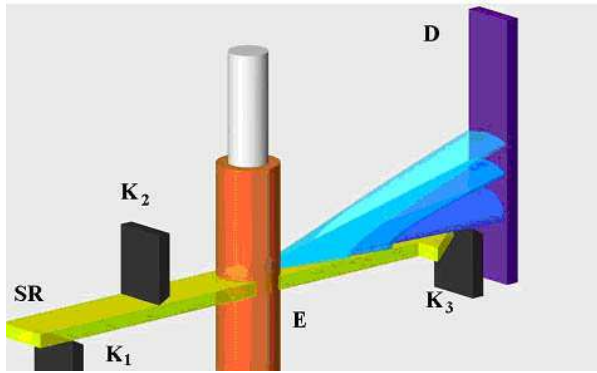}
 \caption{Layout of the SAXS experiment
 using synchrotron radiation.}
\label{f1}
\end{figure}

The experimental set-up is shown in Fig.~1. The SR beam was
collimated with the bottom (K1) and top (K2) edges, and
irradiation zone was formed at the central part of the explosive
charge with 1 mm height and 12--15\,mm width. In front of the
detector (D) the direct beam was closed with another  bottom
edge (K3).  Scattered photons were registered by the detector.
For additional monitoring a part of the direct beam irradiated
the detector through an absorber (1 mm thick copper filter). The
distance between the edges (K1) and (K2) was $\sim\,$200 mm. The
HE charge was at the distance of 700 mm from the edge K2 and at
$\sim\,$640 mm from the edge K3. The distance between the
detector and the edge K3 was $\sim\,$200 mm. Scattered radiation
was registered by the detector DIMEX [3] with the vertical
channel pitch of 0.1 mm. Thus the angular range covered by the
measurements was $\sim\,$0.0068--0.34 degrees. During one flash
of SR the detector recorded all channels (make one frame) and
measured the distribution of SAXS as a function of the angle. As
the detonation front moved along the charge axis with constant
speed 7.6 km/s the series of frames gave time dependence of the
SAXS distribution.

In the investigated HE TNT/RDX 50/50 the duration of the
chemical reaction zone is $\sim\,$0.1\,s [4] and the
measurements of SAXS were performed every 125 ns. VEPP-3 was
operating in two bunch mode for this with bunches rotating in
the opposite points of the orbit. To realize this regime the
problems of positioning, focusing and phase parameters of the
bunches were solved.
\section{Measurement of the dynamics of SAXS from high
explosives with introduced nano-diamonds} The intensity of SAXS
from  mono-dispersed system of  nano-particles can be described
by the following formulae:
\[ I(s)=n(t)\,(\rho-\rho_0)^2F(s),
\]
where  ${\bf s} = {\bf s} - {\bf s}_0$ -- vector of scattering
($|{\bf s}| = 4\pi \sin\theta/\lambda$), $2\theta$ -- scattering
angle, $\lambda$ -- wavelength of radiation), $n(t)$ -- number
of nano-particles of density $\rho$ in a unit volume, $\rho_0$
-- density of the medium, $F(s)$ -- form-factor of an individual
particle.

In the present experiments the maximal registered scattering
angle was  $2\theta_{\rm max}=0.014$, minimum angle was
$2\theta_{\rm min}=0.0006$. Integration in this range of angles
leads to the following formulae for the integral intensity of
SAXS:
\begin{equation}\label{int}
  I_0(\lambda,R)\sim R^{\,6}\,(\rho-\rho_0)^2 N\, .
\end{equation}

Total SAXS intensity can be obtained by integration through all
wavelengths according to the spectrum of irradiation from VEPP-3
storage ring.

In the performed experiments the intensity in (1) is summed from
scattering on atoms (molecules) and scattering on
nano-particles. Analysis shows that  scattering on atoms is
constant and small due to strong dependence on the size in (1).
The registered dependence can be obtained only from relatively
large particles of $\sim\,$2--4 nm size. In the conditions of
the performed experiments the minimal size of detected particles
is equal to  $d_{\rm min}\approx \lambda/\theta_{\rm max}\approx
2.8\,$nm, where $\lambda = 0.04\,$nm is wavelength at the energy
$E=30$ keV.

Maximal size of particles is determined by  minimal registered
scattering angle $\theta_{\rm min}$, that is given by the accuracy of
alignment of the edge that is closing the direct SR beam (around
0.5~mm). For the present experiments
 $d_{\rm max}\approx \lambda/\theta_{\rm min}\approx 70$~íì.

SAXS method does not allow distinguishing of signals from
nano-diamonds and non-diamond carbon forms. As shown in [5]
 fraction of nano-diamond in a solid residue can
 reach 80\% (for initial HE TNT/RDX 50/50).

In all experiments detecting SAXS [2] the signal started from
zero in the zone of maximal compression. One of the possible
reasons of such behavior of SAXS can be very low {\em contrast}
$(\rho_{\rm D}-\rho_1)^2$  of nano-diamond in this moment.
At the density of HE
in the chemical reaction zone of about
$\rho_1 \sim 2.1\,$g/ñm$^3$ the SAXS
signal might not be visible for the detector and only during the
decay of  detonation products, i.e. reduction of the density, it
becomes detectable. Thus it is necessary to know if this method
can distinguish condensed nano-diamonds
 ($\rho_{\rm D}\sim3.5\,$g/cm$^3$) at the
background of shock compressed HE.

In order to answer this question the experiments on SAXS
measurements were performed with detonation of TNT and RDX with
introduced 8\% (weight fraction) of nano-diamonds produced by
NPO ``Altaj''. Such amount of nano-diamonds  corresponds to its
yield at detonation of TNT/RDX 50/50 [5].

The results of the measurements of SAXS are shown in Fig.2. The
initial signal level in TNT and RDX  (curves B and D)
corresponds to 8\% of introduced nano-diamonds. At the moment
when detonation front passes through (zero moment in the figure)
their SAXS signal is reduced proportionally to the change of
{\em contrast}.
For RDX the change of signal have to be equal to
factor of
 $(\rho_{\rm D}-\rho_1)^2/(\rho_{\rm D}-\rho_0)^2 = 1.89$
  where $\rho_{\rm D}\sim3.5\,$g/cm$^3$ is
density of nano-diamonds, $\rho_{0}\sim 1.71\,$g/cm$^3$
is initial density of
RDX and $\rho_{1}\sim2.2\,$g/cm$^3$ is its maximal density.

\begin{figure}
\centering
\includegraphics*[bb=37 10 524 379,width=80mm]{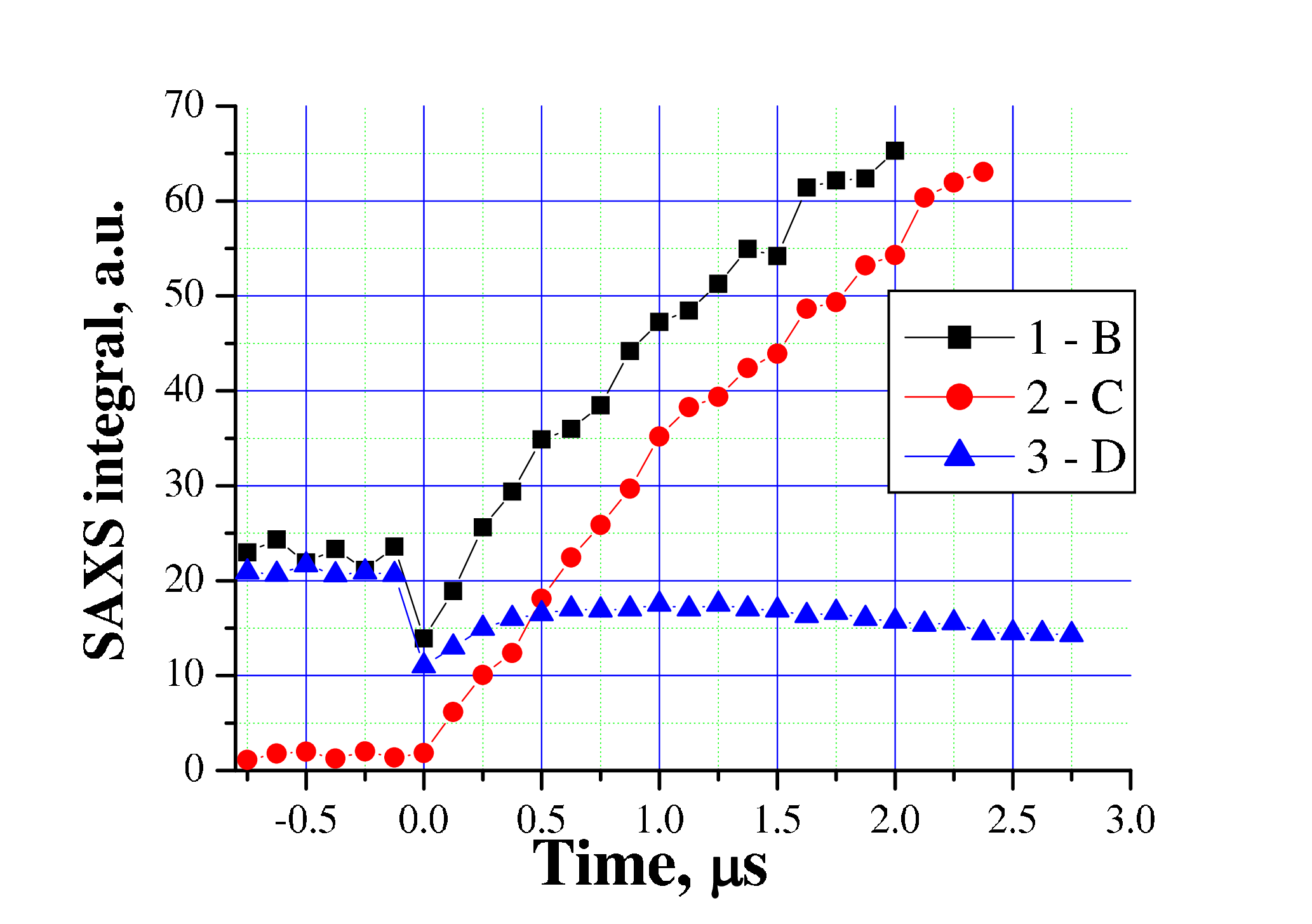}
 \caption{Dependence of integral small-angle scattering
 on time at the detonation of charges of TNT(B),
RDX(D) with introduced 8\% of ultra-dispersed diamonds, and also
of mixture of TNT/RDX 50/50 (C). } \label{f2}
\end{figure}
For TNT the reduction of signal is equal to a factor of
$(\rho_{\rm D}-\rho_1)^2/(\rho_{\rm D}-\rho_0)^2= 1.6$,
 where $\rho_0 = 1.69\,$g/cm$^3$  and  $\rho_1= 2.08\,$g/cm$^3$
 are corresponding densities for TNT. From the plot one
can see that the level of SAXS signal is one order of magnitude
higher than the zero level (noise level). If at detonation of
TNT/RDX 50/50 all nano-diamonds were produced within narrow
chemical reaction zone (in $\sim\,0.1\,\mu$s),
 the jump of SAXS signal to
the same level as in TNT and RDX with introduced nano-diamonds
must be observed in this zone. The behavior of SAXS for TNT/RDX
50/50 (the curve C) demonstrates that in the chemical reaction zone
the signal is practically equal to zero and minimal level of
curves B and D is reached after $\sim\,0.75\,\mu$s
 that significantly
exceeds duration of the chemical reaction. The {\em contrast
factor}
in experiments B and C is practically the same as their decay
dynamics of  the explosion products  differs a little. The SAXS
signal in TNT/RDX 50/50  smoothly grows due to the increase of
the number of nano-particles and their sizes (1), that does not
confirm instant creation of detonation nano-diamonds. Due to
specially taken measures the accuracy of time alignment of
results of different experiments is not worse than 20~ns. Phase
shift between curves B and C at similar level of the signal
significantly exceeds this value.

Behind the detonation front the curve of  SAXS of RDX  with
nano-diamonds passes far below the one of TNT in spite of that
the decay of detonation products in these two cases is
approximately the same. Such behavior of SAXS means destruction
of nano-diamonds in the decaying detonation products of RDX. And
{\em combustion} happens not instantly, but behind the chemical
reaction zone.

\section{Conclusion}
The performed experiments demonstrate that the proposed method
of measurement of SAXS reliably distinguishes exploded
nano-diamonds at the background of substance compressed in a
detonation wave. At detonation of TNT/RDX 50/50 carbon
condensation into nano-diamonds (with sizes $> 2\,$nm)  occurs
behind the chemical reaction zone.

\end{document}